\documentclass[
    ,final            
  ]
  {aipproc}

\layoutstyle{6x9}

\begin{document}

\title{Magnetic fields in neutron stars: A theoretical perspective}

\classification{95.30.Qd; 97.10.Ld; 97.60.Gb; 97.60.Jd}
\keywords      {Magnetic fields --- Magnetohydrodynamics ---
Pulsars --- Neutron stars}

\author{Andreas Reisenegger}{
  address={Departamento de Astronom{\'\i}a y Astrof{\'\i}sica,
  Pontificia Universidad Cat\'olica de Chile,
  Casilla 306, Santiago 22, Chile}
}

\author{Joaqu{\'\i}n P. Prieto}{
  address={Departamento de Astronom{\'\i}a y Astrof{\'\i}sica,
  Pontificia Universidad Cat\'olica de Chile,
  Casilla 306, Santiago 22, Chile}
}

\author{Rafael Benguria}{
  address={Departamento de F{\'\i}sica,
  Pontificia Universidad Cat\'olica de Chile,
  Casilla 306, Santiago 22, Chile}
}
\author{Dong Lai}{
  address={Center for Radiophysics and Space Research, Department of Astronomy,
  Cornell University, Ithaca, NY 14853, USA}
}
\author{Pablo A. Araya}{
  address={Kapteyn Institute, University of Groningen, P. O. Box 800, 9700 AV, Groningen, The Netherlands}
}

\begin{abstract}
We present our view of the main physical ingredients determining
the evolution of neutron star magnetic fields. This includes the
basic properties of neutron star matter, possible scenarios for
the origin of the magnetic field, constraints and mechanisms for
its evolution, and a discussion of our recent work on the Hall
drift.
\end{abstract}

\maketitle


\section{Introduction}
\label{sec:intro}

  In the present conference, Mariano M\'endez reviewed our
  observational knowledge about neutron star magnetic fields,
  while Don Melrose discussed current models of the pulsar
  magnetosphere. Therefore, our review focuses on
  theoretical ideas regarding the magnetic field inside neutron
  stars, paying particular attention to processes by which this
  field might evolve in time. We refer to our previous reviews
  \citep{Reisenegger01a,Reisenegger05} for
  more basic and general discussions of our knowledge of neutron
  star magnetic fields.

  Our discussion starts with an overview of the properties of
  matter in neutron star, which are obviously an essential
  ingredient in any model for the evolution of the magnetic field.
  We continue with a brief discussion of the state of knowledge
  regarding the origin of the magnetic field from the
  main-sequence stage through the hot, protoneutron star phase,
  which sets the initial conditions for the subsequent evolution.
  Then, we quickly present the observational evidence for the evolution
  of the magnetic field, followed by the basic equations that are likely to
  describe this evolution and the main processes involved in it. One
  of these processes, the Hall effect, is singled out for more
  detailed discussion, where we present recent, yet
  unpublished analytic solutions for simple geometries, and
  discuss their implications.

\section{Matter in neutron stars}
\label{sec:matter}

Protoneutron stars are born through gravitational core collapse in
massive stars, which leaves them at a high temperature, $T\sim
10^{11}{\rm K}\sim 10{\rm MeV}$, comparable to the Fermi energies
of their main constituent particles. However, neutrino emission
ensures rapid cooling, so all currently observed neutron stars
(with ages $\sim 10^3{\rm yr}$ or higher) are expected to have
internal temperatures at least two orders of magnitude lower, at
which the matter is highly degenerate and likely close to its
quantum ground state, although small perturbations from the latter
may cause interesting effects
\citep{Reisenegger95,Jones99,Fernandez05}, some of which are
important for the evolution of the magnetic field
\citep{Pethick92,Goldreich92,Jones04} and will be discussed in
what follows.

In this highly degenerate, near-equilibrium state, the properties
of matter are determined by a single parameter, for example
density or pressure. The density inside neutron stars covers many
orders of magnitude, allowing for different regimes in the
properties of matter, as discussed below. For a much more detailed
discussion, see Ref. \citep{Shapiro83}.

\subsection{The crust}
\label{sec:crust}

The neutron star crust is, by definition, the outer part of the
star, where atomic nuclei are present. As in ordinary matter, they
coexist with electrons, which, due to the high density, are too
energetic to be bound to individual nuclei. At $T\sim 10^{10}{\rm
K}$, the nuclei are expected to freeze into a solid, leaving the
electrons as the only moving charge carriers.

Traditionally, it has been assumed that the crust is in its
absolute ground state, in which one particular kind of nucleus,
which minimizes the enthalpy at the local pressure, is present at
any given depth in the star, and these nuclei form a near-perfect
crystal lattice. This picture has been challenged by
\citet{Jones99}, who argued that thermodynamic fluctuations at the
time of freezing would ensure the presence of several kinds of
nuclei at any given pressure, yielding a very impure solid. This
result may be very important for the evolution of the magnetic
field, as it would substantially increase the resistivity of the
neutron star crust, particularly at low temperatures
\citep{Jones04}.

At densities above ``neutron drip'', $\sim 4\times 10^{11}{\rm
g\,cm^{-3}}$, free neutrons also appear within the crust. At a
temperature similar to that for freezing the lattice, these
neutrons are expected to form Cooper pairs and turn superfluid,
largely decoupling dynamically from the rest of the star. This has
been invoked to explain pulsar ``glitches'', quick increases of
the rotation rate of the externally visible parts of a neutron
star, which might be caused by a catastrophic transfer of angular
momentum from the more rapidly rotating superfluid neutrons. Given
that these neutrons are uncharged and interact only weakly with
the electrons, their presence is not likely to be relevant for the
magnetic field evolution.

\subsection{The core}
\label{sec:core}

At $\sim 2\times 10^{14}{\rm g cm^{-3}}$, slightly below standard
nuclear density, the nuclei are expected to loose their individual
identity, so matter at higher densities consists of a liquid-like
mix of neutrons ($n$), protons ($p$), and electrons ($e^-$), which
at progressively higher densities are joined by additional
particles, such as muons, hyperons, and mesons. Chemical
equilibrium among all these particles is established by weak
interactions such as the neutron beta decay ($n\to p+e^-+\bar\nu$)
and inverse beta decay ($p+e^-\to n+\nu$), in most cases involving
the emission of neutrinos ($\nu$) or antineutrinos ($\bar\nu$),
which easily leave the star. In this phase, the strong Pauli
blocking of final states reduces the cross sections for, e.g.,
electron-proton collisions, yielding a very high electrical
conductivity \citep{Baym69}. Here, all particles are movable
(within constraints to be discussed below), making the evolution
of the magnetic field much more complicated.

Models predict that the strongly interacting particles in this
phase ($n$, $p$, possibly hyperons) again form Cooper pairs and
enter a superfluid state, in which the neutron vorticity is
concentrated in quantized vortex lines of microscopic thickness,
much smaller than their average spacing, and the magnetic flux may
be similarly concentrated in proton vortices. The transition
temperatures for these superfluid states are highly uncertain.
Their effect on the thermal evolution of the neutron star,
although potentially quite important, is not obviously shown by
the observations of cooling, young neutron stars (e.g., Fig. 1 of
Ref. \citep{Yakovlev04}), although better fits to the (quite
uncertain) data have been obtained by allowing for neutron stars
with a range of superfluid properties (with mass as the
controlling, free parameter) \citep{Yakovlev04,Page04}. On the
other hand, the core superfluids are not expected to be as easily
decoupled as the crustal neutron superfluid, and therefore play no
role in glitches. Therefore, we consider it safe to say that there
is no strong evidence for the existence of superfluids in the
neutron star core. This conclusion, in addition to the fact that
superfluids introduce a new element of uncertainty in the already
highly complicated models of magnetic field evolution, motivate us
to ignore it in most of what follows, although it has been
discussed in detail by other authors \citep{Mendell98,Ruderman04}.

\section{Origin}
\label{sec:origin}

The range of observed surface magnetic field strengths in
degenerate stars is quite large, from below $10^4{\rm G}$ to
$B_\mathrm{max}\sim 10^9{\rm G}$ on white dwarfs, and from
$10^{11}{\rm G}$ up to $B_\mathrm{max}\sim 10^{15}{\rm G}$ for the
dipole fields on young neutron stars, including classical radio
pulsars, anomalous X-ray pulsars, and soft gamma-ray repeaters
(the field on millisecond pulsars may have been decreased by
processes related to accretion, and not be related to the original
field). The measurable dynamic range on massive main sequence
stars is much smaller, but it is suggestive that the largest
inferred magnetic fluxes in all three kinds of objects are
similar, $\Phi=\pi R^2B_\mathrm{max}\sim 3\times 10^{27}{\rm
G\,cm^2}$, where $R$ is the stellar radius \citep{Reisenegger01a}.
This implies that, in principle, flux freezing from the main
sequence is enough, and no additional magnetic field generation
mechanisms need to be invoked to explain even the strongest fields
observed on the compact stars.

On the other hand, stellar evolution, specially in the massive
stars that eventually become neutron stars, is quite eventful, so
the field geometry (and perhaps its strength) is likely to become
modified along this evolution, possibly coupled to the evolution
of the angular momentum. In particular, the vigorous convection
occurring in a newborn, perhaps rapidly rotating, protoneutron
star is an ideal setting for a dynamo to operate, which might
increase even a fairly weak initial field up to $\sim 10^{15}{\rm
G}$ \citep{Thompson93}.

On the other hand, thermomagnetic instabilities driven by the heat
flow through the neutron star crust
\citep{Urpin80,Blandford83,Wiebicke96} can at most account for
pulsar-like field strengths $\sim 10^{12}{\rm G}$, making them
much less likely to be relevant.

Thus, although no detailed scenario has been convincingly worked
out, it appears likely that a strong magnetic field, permeating
much of the core of the star, is present at least from the hot,
protoneutron star phase onwards.

Regarding its geometrical configuration, recent
magnetohydrodynamic (MHD) simulations by \citet{Braithwaite04}
show that, in a stably stratified star, a complicated, random,
initial field generally evolves on an Alfv\'en-like time scale to
a relatively simple, roughly axisymmetric, large-scale
configuration containing a toroidal and a poloidal component of
comparable strength, both of which are required in order to
stabilize each other. Only the poloidal component crosses the
surface of the star, yielding an external field that is roughly
dipolar, though perhaps somewhat offset from the center, as is
observed in magnetic A stars and white dwarfs.

We should note, however, that both A stars and white dwarfs are of
essentially uniform composition (except for the discontinuity
between the hydrogen-burning, convective core and the chemically
pristine radiative envelope in A stars), so they are stably
stratified only through the entropy gradient in their radiative
regions. This means that, if stable stratification is an essential
ingredient in stabilizing the configurations found in Ref.
\citep{Braithwaite04}, their lifetime would be limited by the
exchange of entropy between regions of different field strengths
\citep{McGregor03}. If, as seems to be the case, this time is
shorter than the lifetime of the stars, the field could be
confined only by compositional gradients (discussed below for the
case of neutron stars), as in the core-crust transition of upper
main-sequence stars.

In any case, it appears plausible that fields of configurations
similar to those of Ref. \citep{Braithwaite04}, with poloidal
field strengths $\sim 10^{11-15}{\rm G}$, and perhaps with their
toroidal component enhanced by the initial differential rotation
\citep{Thompson01}, could be present in early neutron stars.

\section{Evidence for magnetic field evolution}
\label{sec:evidence}

Since the earliest days of the study of pulsars, it has been
claimed \citep{Gunn70,Narayan90} that the distribution of these
objects on the period ($P$) - period derivative ($\dot P$) plane
(sometimes combined with other variables, such as their position
with respect to the Galactic plane, their proper motion, and their
luminosity) would indicate a magnetic field decay on a time scale
comparable to a pulsar lifetime. A safe proof or disproof of such
statements is very difficult to make, since various selection
effects are important, but recent work has generally not confirmed
them \citep{Bhattacharya92,Regimbau01}. Nevertheless, in the past
it provided the motivation for a fair amount of theoretical work
trying to explain the claimed decay.

Another weak argument for magnetic field evolution is the fact
that the braking indices $n\equiv\ddot\Omega\Omega/\dot\Omega^2$
measured in very young pulsars have always turned out to be
smaller than the value ($n=3$) expected from a magnetic dipole
torque \citep{Shapiro83} with a constant dipole moment. This may
indicate that the magnetic field increases in very young pulsars,
or else that the dipole formula is not completely adequate (and
therefore does not give a very reliable estimate of the dipole
field strength \citep{Melatos97}).

A stronger, ``classical'' argument for field decay is the
observation that old neutron stars, such as found in low-mass
X-ray binaries and millisecond pulsars, tend to have much weaker
fields ($\sim 10^{8-9}{\rm G}$) than the younger classical pulsars
and high-mass X-ray binaries ($\sim 10^{12}{\rm G}$). It is still
not clear if this is spontaneous decay happening over the long
lifetime of these objects, or whether it is due to the accretion
process diamagnetically screening the field
\citep{Romani93,Cumming01} or increasing the resistivity to cause
a faster decay. In all these cases, it seems surprising that the
field in all millisecond pulsars appears to reach a bottom value
$\sim 10^8{\rm G}$ and does not decay beyond that. A possible
explanation might be that the original field of the star is
completely screened or lost from the star, whereas new magnetic
flux is carried onto the neutron star by the accretion flow,
limited by the condition that the fluid stresses be strong enough
to compress the field onto the star, which yields a maximum field
strength of about the right order of magnitude.

Finally, but most interestingly, soft gamma-ray repeaters and
anomalous X-ray pulsars appear to radiate substantially more power
than available from their rotational energy loss. At the same
time, their dipole field inferred from the observed torque is
stronger than in all other neutron stars, $\sim 10^{14-15}{\rm G}$
(although a couple of pulsars with $B\sim 10^{14}{\rm G}$ are also
known \citep{Camilo00,Morris02,McLaughlin03}). This makes it
natural to accept the argument of \citet{Thompson96} that these
sources are in fact ``magnetars'', powered by the dissipation of
their magnetic energy (which requires an rms internal field even
stronger, though not by very much, than the inferred dipole
field). In addition, torque changes of both signs, associated with
outbursts from these sources, also argue for a changing magnetic
field structure. These are, at the moment, the only objects in
which there is strong evidence for spontaneous evolution of the
magnetic field.

\section{A physical model}

As explained above, a neutron star core contains a number of
mobile particle species. After an initial transient, following the
formation of the neutron star, on which all sound and Alfv\'en
waves are damped, the evolution of the magnetic field should be
slow enough to make the inertia of the particles negligible,
leading to the equation of diffusive motion for the particles of
each species $i$,

\begin{equation}
0=-\nabla\mu_i-m_i^*\nabla\psi+q_i\left(\vec E+{\vec v_i\over
c}\times\vec B\right)-\sum_j \gamma_{ij}n_j(\vec v_i-\vec v_j),
\label{diffusion}
\end{equation}

\noindent where $\mu_i$, $m_i^*$, $q_i$, $\vec v_i$ are their
chemical potential (Fermi energy), effective mass (including
relativistic corrections due to random motions and interactions),
electric charge, and mean velocity, $\psi$ is the gravitational
potential, $\vec E$ and $\vec B$ are the electric and magnetic
fields, and the last term represents the momentum transfer due to
collisions with all other particle species $j$, each with number
density $n_j$. The collisional coupling strengths are
parameterized by the symmetric matrix $\gamma_{ij}$, whose
coefficients generally depend on position.

The collision terms have two effects in the context of the
evolution of the magnetic field:

\begin{itemize}
\item[(a)] They damp the relative motion of positive and negative
charge carriers, leading to a resistive diffusion of the magnetic
field, opposed by the induced electric field. For a magnetic field
of spatial scales comparable to the neutron star radius, the
induction is large and the resistance is small, as in most
astrophysical settings, so this process is slow and unlikely to
have any effect over the lifetime of any neutron star
\citep{Baym69}. It may, however, be important if the magnetic
field is created within a thin surface layer (which does not fit
with the likely formation scenarios we described) or if
small-scale structure is created by other processes, as discussed
below.

\item[(b)] Collisions tend to keep the different particle species
moving with similar velocities, as in the standard MHD
approximation. We now proceed to analyze this type of motion.
\end{itemize}

Adding the forces given by eq. \eqref{diffusion} over all
particles (of all species) within a volume containing a unit total
mass and no net charge, one obtains the equation of MHD
equilibrium,

\begin{equation}
0=-{\nabla P\over\rho}-\nabla\psi+{\vec j\times\vec B\over\rho c},
\label{hydrostatic}
\end{equation}

\noindent where the mass density $\rho=\sum_i n_i m_i^*$, the
current density $\vec j=\sum_i n_i q_i \vec
v_i=(c/4\pi)\nabla\times\vec B$, and the pressure gradient term
was obtained from the zero-temperature Gibbs-Duhem relation,
$dP=\sum_i n_i d\mu_i$. Taking the curl of eq.
\eqref{hydrostatic}, we obtain

\begin{equation}
{\nabla P\times\nabla\rho\over\rho^2}=\nabla\times\left(\vec
j\times\vec B\over\rho c\right).
\label{curl}
\end{equation}

The right-hand side is generally non-zero, therefore equilibrium
is only possible if the pressure and density gradients appearing
on the left-hand side are not parallel. This is not possible in
cold matter in chemical equilibrium, since then the density is a
unique function of pressure (because all the chemical abundances
are also determined by the latter). Therefore, the magnetic field
generally perturbs the chemical equilibrium, causing a tiny
misalignment between the pressure and density gradients.

A magnetic force distribution that has a horizontal curl component
induces compensated up and downward motions of the fluid in
different regions, producing the mentioned misalignment of the
density and pressure gradients, which chokes the motion. Only the
horizontal fluid motions produced by a magnetic force density with
a purely vertical curl can proceed essentially unimpeded. This is
a manifestation of the stable stratification of the neutron star
matter \citep{Pethick92,Reisenegger92,Reisenegger01b}, which
prevents even a magnetar-strength field from being transported
vertically through the star. The only ways by which this
constraint can be circumvented are \citep{Pethick92,Goldreich92}:

\begin{itemize}
\item[(a)] elimination of the induced chemical imbalance by weak
interaction processes, which is most effective at high
temperatures, where reactions are fast, and

\item[(b)] relative motion of different particle species.
\end{itemize}

We now proceed to discuss each of these processes.

\subsection{Bulk motion facilitated by weak interactions}
\label{sec:bulk}

As described above, the Lorentz force produces non-barotropic
perturbations to the pressure, of magnitude $\delta P\sim
B^2/8\pi\sim\sum n_i\delta\mu_i$, where the relation to the
chemical potential perturbations again comes from the Gibbs-Duhem
relation. If, for definiteness, we consider the simplest possible
neutron star matter, composed of neutrons, protons, and electrons
(the latter two related by the condition of charge neutrality,
$n_p=n_e\equiv n_c$), the total chemical imbalance is
\begin{equation}
\Delta\mu\equiv|\mu_p+\mu_e-\mu_n|\sim B^2/(8\pi n_c)\sim
3B_{15}^2~\mathrm{keV}. \label{imbalance}
\end{equation}
where $B_{15}\equiv B/(10^{15}{\rm G})$. An asymmetry in the weak
interaction rates tends to reduce this imbalance. In what follows,
we assume that the dominant process are ``modified Urca
reactions'' without Cooper pairing, which give a reasonable fit to
the early cooling of neutron stars (see Fig. 1 in Ref.
\citep{Yakovlev04}) and possibly to the late, ``rotochemical''
reheating of millisecond pulsars \citep{Fernandez05}. As long as
$T\gg\Delta\mu$ \citep{Reisenegger95}, the imbalance decays
exponentially, with time constant $t_\mathrm{mU}\sim
0.5/T_9^6~\mathrm{yr}$, essentially the cooling time of the star,
which is highly sensitive to temperature ($T=T_9\times
10^9~\mathrm{K}\approx T_9\times 86~\mathrm{keV}$). In this time,
the Lorentz force makes the fluid move a small fraction,
$\sim\Delta\mu/\mu_e\sim 2\times 10^{-5}B_{15}^2$, of the stellar
radius, so the total decay time of the field is
\begin{equation}
t_\mathrm{decay}\sim {\mu_e\over\Delta\mu}t_\mathrm{mU}\sim
{3\times 10^4\over B_{15}^2T_9^6}~\mathrm{yr}, \label{bulkdecay}
\end{equation}
in principle quite short for strong fields and high temperatures.
However, it is important to note that it is longer than the
cooling time by the generally quite considerable ratio
$\mu_e/\Delta\mu$. Thus, unless the magnetic field is in the
vicinity of $\sim 10^{17}~\mathrm{G}$ (making
$\Delta\mu\sim\mu_e$), a heat source is required to prevent
passive cooling of the neutron star from ``freezing'' the magnetic
field. The most obvious energy source is the magnetic field itself
\citep{Thompson96}, whose energy in this scenario is released by
the weak interactions. We note, however, that, in the absence of a
chemical imbalance, the neutrino emission produced by these weak
interaction leads to the cooling of the star. It is balanced by
the heat release only if the chemical imbalance is fairly large,
$\Delta\mu\approx 5.5 T$ \citep{Fernandez05} or, equivalently,
$B_{15}\approx 13 T_9^{1/2}$. A highly magnetized neutron star
would be born at a high temperature and quickly cool down until
this condition is satisfied, after which the temperature would be
stabilized (at $T_9\sim 0.2[10^4~\mathrm{yr}/t]^{1/7}$) by the
energy injected through the decaying magnetic field,\footnote{The
coefficient is slightly higher than found by \citet{Thompson96},
who first obtained essentially the same result.} $B_{15}\sim
6.5(10^4~\mathrm{yr}/t)^{1/14}$. At this fairly high temperature,
the collision rates are high, thus the single-fluid MHD
approximation is probably adequate.

\subsection{Relative motion of charged and neutral particles}
\label{sec:relative}

The process just described is not relevant for stars with internal
fields substantially lower than $5\times 10^{15}~\mathrm{G}$,
which would enter the photon-cooling epoch \citep{Yakovlev04}
before the magnetic field has substantially decayed or contributed
to any reheating. For these, the only possibility to have magnetic
field decay relies on relative motion of charged particles. We
consider a slight extension of the model of $npe$ matter proposed
by \citet{Goldreich92}, in which different linear combinations of
the eqs. \eqref{diffusion} for the three particle species,
together with the induction equation $\partial\vec B/\partial
t=-c\nabla\times\vec E$, yield
\begin{equation}
{\partial\vec B\over\partial t}=\nabla\times\left[(\vec v_n+\vec
v_A+\vec v_H)\times\vec B\right]-\nabla\times\left(c\vec
j\over\sigma\right)-{c\over
2e}\nabla\left(\gamma_{en}-\gamma_{pn}\over\gamma_{en}+\gamma_{pn}\right)\times\nabla(\mu_p
+\mu_e). \label{evolution}
\end{equation}
In the absence of the last two (resistive and battery) terms,
which in realistic conditions are generally small, the magnetic
field is advected by a combination of three velocities:
\begin{itemize}
\item[(a)] the neutron velocity $\vec v_n$, corresponding to the
bulk motions considered in the previous subsection; \item[(b)] the
ambipolar diffusion velocity,
\begin{equation}
\vec v_A\equiv{\gamma_{pn}(\vec v_p-\vec v_n)+\gamma_{en}(\vec
v_e-\vec v_n)\over \gamma_{pn}+\gamma_{en}}={\vec j\times\vec
B/(n_cc)-\nabla(\Delta\mu)\over
(n_n+n_c)(\gamma_{pn}+\gamma_{en})}, \label{ambipolar}
\end{equation}
representing a relative motion of the charged particles with
respect to the neutrons, driven by the Lorentz force, and
potentially choked by the gradient it induces in the chemical
potential imbalance; and \item[(c)] the Hall drift velocity,
\begin{equation}
\vec v_H\equiv
{\gamma_{en}-\gamma_{pn}\over\gamma_{en}+\gamma_{pn}}(\vec
v_p-\vec
v_e)={\gamma_{en}-\gamma_{pn}\over\gamma_{en}+\gamma_{pn}}{\nabla\times\vec
B\over 4\pi n_cc}, \label{Hallvel}
\end{equation}
\end{itemize}
due to the relative motion of protons and electrons and
proportional to the current density.

The chemical imbalance induced by ambipolar diffusion can in
principle be eliminated by weak interactions, on time scales
similar to those of the bulk motions, which are short only at high
temperatures, at which ambipolar diffusion is strongly suppressed
by collisions. The regime in which it may be important is at lower
temperatures, where collisions are less constraining, but weak
interactions are strongly suppressed. In this case, only the
solenoidal mode of ambipolar diffusion (driven by the finite-curl,
zero-divergence part of the Lorentz force \citep{Goldreich92}) can
proceed (and even this only in the simplified case of a uniform
charged fluid with $n_p=n_e$, without the stabilizing effect of
nonuniform relative particle abundances if more charge carriers
are present).

The Hall drift is of a very different character, since the drift
velocity is directly related to the magnetic field, and
independent of any arising fluid forces. In a solid medium such as
the neutron star crust, where the electrons are the only free
charge carrier, it is the only active process besides resistive
diffusion. This regime has been studied by several authors and is
discussed in the following section. The more complicated regime of
the fluid core, in which the Hall effect acts in conjunction with
(or in opposition to) a highly constrained ambipolar diffusion, is
largely an open question \citep{Arras04}.

\section{Hall drift}
\label{sec:hall}

In the neutron star crust or another solid, conducting medium, the
only moving charges are the electrons, and the evolution of the
magnetic field is described by the ``Hall equation'',
\begin{equation}
{\partial\vec B\over\partial t}=\nabla\times\left[\left(-{c\over
4\pi n_ee}\nabla\times\vec B\right)\times\vec B+{c^2\over
4\pi\sigma}\nabla\times\vec B\right], \label{Halleq}
\end{equation}
where the two terms on the right-hand side correspond to the Hall
drift and resistive diffusion, respectively, whose relative
importance in the neutron star crust is still a matter of
controversy \citep{Cumming04,Jones04}.

Here, we discuss some physical issues that are likely to determine
the evolution of magnetic field under the Hall effect, without
attempting to cover the many simulations recently performed to
address how specific magnetic field configurations might decay in
real neutron stars \citep{Rheinhardt04,Hollerbach04}.

\citet{Goldreich92} focused on the more interesting case in which,
on large scales (comparable to the crust thickness) the Hall
effect is dominant, and argued, by analogy with the Euler equation
of fluid dynamics, that the nonlinear Hall term may give rise to a
turbulent cascade to small scales. Due to the presence of stable,
linear modes, this turbulence would be ``weak'', with an energy
transfer time generally longer than the typical oscillation period
on a given scale, resulting in a power spectrum $\propto k^{-2}$,
slightly different from the Kolmogorov spectrum $\propto k^{-5/3}$
of fluid turbulence. At small scales, the magnetic energy would
finally be dissipated by resistivity. This nonlinear evolution was
simulated by \citet{Biskamp99}, who found an energy cascade to
small spatial scales, but with an even steeper spectrum than
predicted, $\propto k^{-7/3}$.

A complementary approach is to study analytic solutions of the
Hall equation (with or without the resistive term), which has been
done by \citet{Vainshtein00}, \citet{Cumming04}, and by our group
\citep{Araya02,Prieto04}. In the latter, which extends and
generalizes the work of Ref. \citep{Vainshtein00}, we first
considered a purely toroidal field, $\vec B={\cal
B}(R,z,t)\nabla\phi$, where $R,\phi,z$ are the standard
cylindrical coordinates. When evolved by the Hall equation, the
field remains toroidal. In order to describe its evolution, it
becomes convenient to introduce a new coordinate $\chi\equiv
c/[4\pi en_e(R,z)R^2]$. Surfaces of constant $\chi$ (hereafter,
$\chi$-surfaces) are toroids contained inside the star. A
complementary coordinate $s$ can be defined on each $\chi$-surface
by $\partial/\partial s\equiv
-R^2\nabla\phi\times\nabla\chi\cdot\nabla$, in terms of which the
Hall equation (now neglecting the resistive term) reduces to the
Burgers equation,
\begin{equation}
{\partial{\cal B}\over\partial t}+{\cal B}{\partial{\cal
B}\over\partial s}=0, \label{Burgers}
\end{equation}
with the well-known, implicit solution ${\cal B}=f(s-{\cal B}t)$.
Each value of ${\cal B}$ is carried around the corresponding
$\chi$-surface with a velocity proportional to ${\cal B}$,
developing discontinuities at the (comoving) points where
${\partial{\cal B}/\partial s}$ is large. There, the resistive
term increases and tends to smooth the discontinuity. Therefore,
magnetic energy is dissipated at the (relatively fast) rate as it
is fed into the discontinuity by the Hall drift. Eventually, on
the characteristic time scale of the Hall drift, ${\cal B}$
becomes uniform on each $\chi$-surface, so the resulting magnetic
field $\vec B={\cal B}(\chi)\nabla\phi$ only evolves on the
(assumed) much longer resistive time scale.

However, further analysis \citep{Prieto04} shows that this field
is unstable to small, poloidal perturbations, which grow when
different segments of the poloidal field lines, crossing different
$\chi$-surfaces, are carried along at different speeds by the Hall
drift velocity associated to the toroidal field. So, the question
arises: Are there any non-trivial configurations which are not
only static, but also stable, under the Hall effect?

A likely important clue is conservation of magnetic helicity,
\begin{equation}
{\partial\over\partial t}(\vec A\cdot\vec
B)+\nabla\cdot(c\varphi\vec B+c\vec E\times\vec A)=-2c\vec
E\cdot\vec B. \label{helicity}
\end{equation}
This equation, derived directly from Maxwell's equations (where
$\varphi$ and $\vec A$ are the standard electromagnetic scalar and
vector potentials), shows that the volume integral of the
``magnetic helicity density'' $\vec A\cdot\vec B$ is conserved if
$\vec E\cdot\vec B\equiv 0$, which is the case whenever the
resistivity is zero, so the generalized Ohm's law reduces to $\vec
E+\vec v\times B/c=0$, regardless of the specific form of the
velocity field $\vec v$. In particular, the stable MHD equilibria
found by \citet{Braithwaite04} most likely minimize energy at a
given magnetic helicity, and the Hall effect also conserves this
quantity.

Since, dimensionally, the magnetic helicity density is $\sim B^2L$
(where $B$ and $L$ are a characteristic magnetic field strength
and length scale), whereas the magnetic energy density is $\sim
B^2$, the helicity tends to reside on large scales, making it much
more difficult to dissipate than energy. This could make strongly
helical configurations stable and prevent them from decaying under
any of the processes considered (aside from the extremely slow
resistive diffusion). We are currently engaged in identifying such
configurations.

\section{Conclusions}
\label{sec:conclusions}

The evolution of the magnetic field in neutron stars is a very
challenging subject, with strong observational clues and involving
complex physics. Perhaps surprisingly (given the extreme
conditions of density, gravity, and field strength), it is likely
to involve many of the same processes also emerging in the
evolution of magnetic fields in other contexts, from plasma
physics laboratories to galaxies, so continued cross-feeding of
insights should be beneficial both to the study of neutron stars
and of other systems.

\begin{theacknowledgments}
  We thank Peter Goldreich and Chris Thompson for highly illuminating discussions.
  Financial support from FONDECYT (Chile) through Regular Grants 1020840 and 1020844
  and International Cooperation Grant 7020840 is gratefully acknowledged.
\end{theacknowledgments}



\bibliographystyle{aipproc}   

\end{document}